\documentclass[11pt,twoside]{article}

\usepackage{asp2006}
\usepackage{epsf}
\usepackage{psfig}
\usepackage{lscape}

\begin{document}
\title{Parsec Scale Properties of Brightest Cluster Galaxies}   
\author{G. Giovannini, E. Liuzzo, M. Giroletti}   
\affil{Dipartimento di Astronomia, Universita' di Bologna, via Ranzani 1
40127 Bologna, Italy \\
and \\
Istituto di Radioastronomia - INAF, via Gobetti 101
40129 Bologna, Italy}    

\begin{abstract}

We present new VLBI observations of 
Brightest Cluster Galaxies in eight nearby Abell clusters. 
These data show a possible difference
between Brightest Cluster Galaxies in cool core clusters (two-sided pc scale 
jets) and in non cool core
clusters (one-sided pc scale jets). We suggest that this difference could be 
due to the jet interaction with the surrounding medium. More data are necessary
to discuss if pc-scale properties of Brightest Cluster Galaxies are 
influenced by their peculiar morphology and position in the center of rich 
clusters of galaxies.

\end{abstract}

\section{Introduction}   

Brightest Cluster Galaxies (BCGs) are a unique class of objects (e.g.
Lin \& Mohr 2004).
They tend to lie very close to peaks of the clusters X-ray emission
and in the velocity space they sit near the cluster rest frame.
These galaxies are the most
luminous and massive galaxies in the universe. The BCGs are variously
classified as giant Ellipticals, D galaxies, or cD galaxies because
of the extended envelope of excess light.
The optical morphology often shows evidence of past or recent
galaxy mergers (e.g. multiple nuclei).
All these properties indicate that they may have a quite unusual formation
history compared to other E galaxies.
These galaxies are intimately related to the collapse and formation of the
cluster: recent models suggest that BCGs must have an earlier origin and
that galaxy merging within the cluster during collapse in cosmological
hierarchy is a possible alternative.

In the radio band BCGs are more likely to host radio-loud AGN than other
galaxies of the same mass (Best et al. 2006) and very often their radio
morphology shows evidence
of a strong interaction with the surrounding medium as a distorted tailed
structure (WAT sources), strong confinement, buoyancy effects and so on.
The radio morphology is very
various: some BCGs show a standard tailed structure (WAT) very extended on
the kpc scale as 3C465 in A2634 (e.g. Eilek et al. 1984), or with a small 
size as NGC 4874 in the Coma
cluster (Feretti \& Giovannini 1985); in other
cases we have diffuse and amorphous sources, either extended (e.g. 3C 84
in Perseus) or with very small size (e.g. A154, Feretti \& Giovannini 1994).

The recent result that in every/most of cooling core clusters is present
an active radio BCG (Eilek \& Owen, 2006)
suggests that the radio loud AGN could be the origin  of
the energy necessary to arrest or slow down the cooling process as shown
by the presence of cavity in the radio emitting gas coincident with
the presence of radio lobes (see e.g. Dunn \& Fabian 2008 
and references there in).

On the parsec scale BCGs are not yet well studied as a class of source.
Only a few of them have been observed because are well known radio galaxies.
In some cases they look as normal FR I radio galaxies with relativistic
collimated jets often one-sided because of Doppler boosting effects (e.g. 3C465
in A 2634, and 0836+29 in A690; Venturi et al. 1995).
However there are also cases where two-sided symmetric jets are present in 
VLBI images, 
and it is not clear
if they are highly relativistics or not (e.g. 3C338 in A2199, Gentile et
al. 2007).

\section{Observations and Results}

We started a PhD project to observe with VLBA (NRAO) a complete sample of
BCGs in Abell
Clusters selecting nearby clusters (Distance Class lower than 3) with a
Declination larger than 0$^\circ$. All clusters have been included with
no selection on the cluster conditions (e.g cooling) and no selection on the
BCG radio power. The sample is presented in Table 1, where (1) in the notes
column indicates clusters where we observed the BCG with VLBA observations
at 6 cm, (2) means that we asked for VLBA observing time, and (3) clusters
where the BCG is a well known radio galaxy with published VLBI data.

\begin{table}[!ht]
\caption{The Cluster Sample}
\smallskip
\begin{center}
{\small
\begin{tabular}{ccccccc}
\tableline
\noalign{\smallskip}
Abell Cluster & z & DISTCL & RA$_{J2000}$ & DEC$_{J2000}$ & notes & Cooling \\
\noalign{\smallskip}
\tableline
\noalign{\smallskip}
 262& 0.0161& 1&  01 52 50&  36 08   & (1)& y\\
 347& 0.0187& 1&  02 25 50&  41 52   & (2)& y\\
 400& 0.0232& 1&  02 57 38&  06 02   & (2)& n\\
 407& 0.0470& 2&  03 01 43&  35 49   & (2)& y\\
 426& 0.0183& 0&  03 18 36&  41 31   & (3)& y\\
 539& 0.0205& 2&  05 16 35&  06 27   & (2)& n\\
 569& 0.0196& 1&  07 09 10&  48 37   & (1)& n\\
 576& 0.0381& 2&  07 21 24&  55 44   & (2)& n\\
 779& 0.0226& 1&  09 19 50&  33 46   & (2)& n\\
1185& 0.0304& 2&  11 10 47&  28 40   & (2)& n\\
1213& 0.0468& 2&  11 16 29&  29 15   & (2)& n\\
1228& 0.0350& 1&  11 21 29&  34 19   & (2)& n\\
1314& 0.0341& 1&  11 34 48&  49 02   & (1)& n\\
1367& 0.0215& 1&  11 44 29&  19 50   & (1)& n\\
1656& 0.0232& 1&  12 59 48&  27 59   & (1)& n\\
2147& 0.0356& 1&  16 02 17&  15 53   & (1)& n\\
2151& 0.0371& 1&  16 05 15&  17 45   & (1)& n\\
2152& 0.0374& 1&  16 05 22&  16 27   & (1)& y\\
2162& 0.0320& 1&  16 12 30&  29 32   & (2)& n\\
2197& 0.0303& 1&  16 28 10&  40 54   & (2)& y\\
2199& 0.0303& 1&  16 28 36&  39 31   & (3)& y\\
2634& 0.0312& 1&  23 38 18&  27 01   & (3)& n\\
2666& 0.0265& 1&  23 50 56&  27 08   & (2)& n\\
\noalign{\smallskip}
\tableline
\end{tabular}
}
\end{center}
\end{table}

\begin{figure}[!ht]
\plottwo{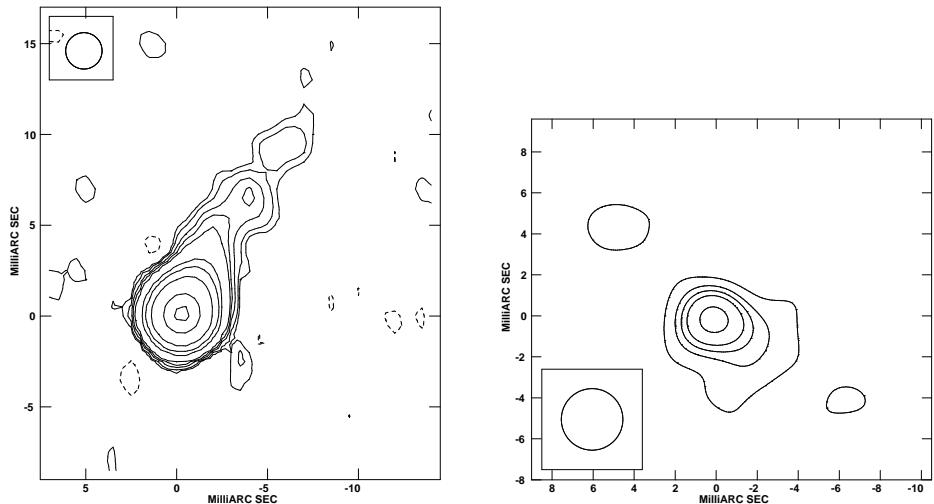}{ggfig1b}
\caption{1a (left): VLBA image at 5 GHz of NGC2329, the BCG in A569. Levs 
are: -0.2 0.2 0.3 0.5 0.7 1 3 5 10 30 50 mJy/beam. 
1b (right): VLBA image at 5 GHz of NGC4874, the BCG in A1656 (Coma 
cluster). Levs are: -0.2 0.15 0.3 0.4 0.5 0.7 mJy/beam.}
\end{figure} 

\begin{table}[!ht]
\caption{Results}
\smallskip
\begin{center}
{\small
\begin{tabular}{ccccc}
\tableline
\noalign{\smallskip}
Cluster Name & cool core & BCG & Large Scale & VLBI \\
\noalign{\smallskip}
\tableline
\noalign{\smallskip}
A569  & N & NGC2329& WAT                 & one-sided \\
A1314 & N & IC712  & WAT small           & N.D.      \\
      & N & IC708  & WAT                 & one-sided \\
A1367 & N & NGC3842& WAT small           & N.D.      \\
      & N & 3C264  & HT                  & one-sided \\
A1656 & N & NGC4874& WAT small           & one-sided \\
A2151 & N & NGC6041& WAT small           & core      \\
A2634 & N & 3C465  & WAT                 & one-sided \\
      &   &        &                     &           \\
A262  & Y & NGC708 & Relaxed Double      & N.D.      \\
A426  & Y & 3C84   & Compact core + Halo & two-sided \\
A2199 & Y & 3C338  & Double - Restarted  & two-sided \\

\noalign{\smallskip}
\tableline
\end{tabular}
}
\end{center}
\end{table}

We present here preliminary results from our and literature data. We observed
each source for about 3 hrs to assure a good uv-coverage and a low
noise level. The resolution of final maps is typically 2.8 $\times$ 1.5 mas
and the noise level is $\sim$ 0.05 mJy/beam.

Among observed cluters mostly detected sources show a one-sided structure,
 but we did not detect three sources despite of the high sensitivity of
our images. In particular we did not detect
the BCG in A262 identified with a B2 radio galaxy (NGC 708; Parma et al. 1986)
suggesting that at present the core is not in a very active phase.

In Table 2 we show the list of observed clusters and of
the few clusters with published data. We have separated non relaxed clusters
and relaxed clusters with a cool core. The number of clusters with radio
data is small, and results are
very preliminary, however we note that in non relaxed clusters the large scale
radio structure suggests a strong interaction with the surrounding medium
(Wide Angle Tail or Head Tail sources) and in most of these sources we
found one-sided parsec scale jets. We interpret this structure as 
due to Doppler boosting effects in relativistic, intrinsically symmetric jets.
The two BCGs in clusters with a
cool core show a two-sided symmetric pc scale structure, and for both sources 
it is not clear if jets are highly relativistic or not (the BCG in A262 has
not been detected in our VLBI images). 

This difference between relaxed and non relaxed clusters could
suggest that in the parsec scale region of cool clusters there is a 
strong interaction between radio jets
and the surrounding medium and jets slow down very soon. 
This suggestion is supported by literature data
on the BCGs of more distant clusters as e.g. 
4C26.42, the BCG in the cool cluster A1795, where VLBI images show a distorted
symmetric structure (Liuzzo et al. in preparation), and 
Hydra A (the BCG of the cool cluster A780), where 
Taylor (1996) suggested that the emission
from the symmetric pc scale jets is more dependent on interactions with the 
surrounding material than on Doppler boosting. 

\acknowledgements 
We thank the organizers of a very interesting meeting.

The National Radio Astronomy Observatory is operated by Associated 
Universities, Inc., under cooperative agreement with the National Science 
Foundation.


\begin{thebibliography}{}
\bibitem[Best et al.(2006)]{bes06}
Best, P. N., Kaiser, C. R., Heckman, T. M., Kauffmann, G. 2006 \mnras ~~368, 67
\bibitem[Dunn \& Fabian(2008)]{dul08}
Dunn, R. J. H.\& Fabian, A. C. 2008 \mnras ~~in press; arXiv:0801.1215
\bibitem[Eilek et al.(1984)]{eil84}
Eilek, J. A., Burns, J. O., Odea, C. P., Owen, F. N. 1984 \apj ~~278, 37
\bibitem[Eilek \& Owen(2006)]{eil06}
Eilek, J. A. \& Owen, F. N. 2006 To appear in proceeding of ``Heating and 
Cooling in Galaxies and Clusters of Galaxies'' (August 2006, MPE/Garching);
arXiv:astro-ph/0612111
\bibitem[Feretti \& Giovannini(1985)]{fer85}
Feretti, L. \& Giovannini, G. 1985 \aap ~~147, 13
\bibitem[Feretti \& Giovannini(1994)]{fer94}
Feretti, L. \& Giovannini, G. 1994 \aap ~~281, 375
\bibitem[Gentile et al.(2007)]{gen07}
Gentile, G. Rodriguez, C. Taylor, G. B. Giovannini, G. Allen, S. W. 
Lane, W. M. Kassim, N. E. 2007 \apj ~~659, 225
\bibitem[Lin \& Mohr(2004)]{lin04}
Lin, Y-T.. \& Mohr, J. J. 2004 \apj ~~617, 879 
\bibitem[Parma et al.(1986)]{par86}
Parma, P. de Ruiter, H.R. Fanti, C. Fanti, R. 1986 \aaps ~~64, 135 
\bibitem[Taylor(1996)]{tay96}
Taylor G.B. 1996 \apj ~~470, 394
\bibitem[Venturi et al.(1995)]{ven95}
Venturi, T. Castaldini, C. Cotton, W. D. Feretti, L. Giovannini, G. Lara, L. 
Marcaide, J. M. Wehrle, A. E. 1995 \apj ~~454, 735

\end{thebibliography}
\end{document}